\documentclass[%
aps,
reprint,
prl,
superscriptaddress,
nofootinbib,
nobibnotes,
showkeys,
floatfix,
]{revtex4-2}

\usepackage{xcolor}
\usepackage{graphicx}
\graphicspath{{fig/}}
\usepackage{amsmath,amssymb,amsfonts,mathrsfs,bm}
\usepackage[%
colorlinks=true,
linkcolor=blue,
citecolor=blue,
]{hyperref}

\usepackage{iftex}
\ifpdftex% Commands and packages specific to pdfLaTeX
    \usepackage{CJKutf8}
    \newcommand{\Ckai}[1]{\begin{CJK*}{UTF8}{gkai}#1\end{CJK*}}
\fi
\ifxetex% Commands and packages specific to XeLaTeX
    \usepackage{xeCJK}
    \newcommand{\Ckai}[1]{\textit{#1}}
\fi

\newcommand{\dif}{\mathrm{d}}

\newcommand{\Msun}{\,M_{\odot}}

\newcommand{\second}{\,\mathrm{s}}
\newcommand{\eV}{\,\mathrm{eV}}

\newcommand{\pc}{\,\mathrm{pc}}
\newcommand{\kpc}{\,\mathrm{kpc}}
\newcommand{\Mpc}{\,\mathrm{Mpc}}
\newcommand{\Gpc}{\,\mathrm{Gpc}}

\newcommand{\muas}{\,\mu\mathrm{as}}

\newcommand{\DM}{\mathrm{DM}}

\newcommand{\jmax}{j_{\max}}
\newcommand{\alphahat}{\hat{\alpha}}
\newcommand{\alphahatpk}{\alphahat_{\mathrm{peak}}}

\newcommand{\rc}{r_{\mathrm{c}}}

\newcommand{\rcS}{r_{\mathrm{cS}}}

\newcommand{\DLS}{D_{\mathrm{LS}}}
\newcommand{\DOL}{D_{\mathrm{OL}}}
\newcommand{\DOS}{D_{\mathrm{OS}}}

\newcommand{\thetacri}{\theta_{\mathrm{cri}}}
\newcommand{\xcri}{x_{\mathrm{cri}}}
\newcommand{\xEin}{x_{\mathrm{Ein}}}
\newcommand{\thetaEin}{\theta_{\mathrm{Ein}}}
\newcommand{\cau}{\mathrm{cau}}
\newcommand{\betacau}{\beta_{\cau}}
\newcommand{\ycau}{y_{\cau}}
\newcommand{\src}{\mathrm{src}}

\newcommand{\evt}{\mathrm{evt}}
\newcommand{\Rsrc}{R_\src}
\newcommand{\abs}{\mathrm{abs}}
\newcommand{\app}{\mathrm{app}}

\begin{document}

\title{%
    Periodic Gravitational Lensing with Oscillating Boson Stars
}

\author{Xing-Yu Yang (\Ckai{杨星宇})}
\email[Corresponding author:~]{xingyuyang@kias.re.kr}
\affiliation{Quantum Universe Center (QUC), Korea Institute for Advanced Study, Seoul 02455, Republic of Korea}

\author{Tan Chen (\Ckai{陈坦})}
\email[Corresponding author:~]{chentan@bnu.edu.cn}
\affiliation{School of Physics and Astronomy, Beijing Normal University, Beijing 100875, China}
\affiliation{Institute for Frontier in Astronomy and Astrophysics, Beijing Normal University, Beijing 102206, China}

\author{Rong-Gen Cai (\Ckai{蔡荣根})}
\email[Corresponding author:~]{caironggen@nbu.edu.cn}
\affiliation{Institute of Fundamental Physics and Quantum Technology, Ningbo University, Ningbo 315211, China}

\begin{abstract}
    We show that oscillating (real-scalar) boson stars generically host an \emph{oscillating radial caustic}. Sources near this caustic cross it every half period, thereby producing periodic caustic-crossing lensing. The resulting observables are phase locked to the lens oscillation: image-pair creation or annihilation, changing image morphology, achromatic photometric spikes, and astrometric motion. This signal provides a distinctive target for time-domain astronomy, and its detection would reveal an intrinsically time-dependent compact dark-sector object. Event-number estimates indicate a measurable discovery space with current astrometric and high-cadence photometric surveys, while null searches would constrain the abundance of such objects as dark matter. The predictions rely only on the dynamics of real-scalar condensates and extend naturally to self-interacting real scalars, including axionlike particles, and to ultralight vector bosons.
\end{abstract}

\maketitle

\textit{Introduction---}%
Boson stars are self-gravitating condensates of bosonic quantum fields whose structure is fixed by quantum pressure, gravity, and possible self-interactions~\cite{Kaup:1968zz}.
They appear across particle-physics scenarios ranging from ultralight axions to Higgs-portal scalars and can form compact dark-sector objects over many mass scales~\cite{Jetzer:1991jr, Schunck:2003kk, Marsh:2015xka, Ferreira:2020fam}.
They also motivate probes of strong-gravity compact objects, black-hole alternatives, and dark matter~\cite{Liebling:2012fv, Croon:2020wpr, Croon:2020ouk, Visinelli:2021uve, Lira:2024cma, Banik:2025xwl}.

For a real-scalar field, the absence of a conserved U(1) charge makes the stress energy and spacetime oscillate coherently~\cite{Seidel:1991zh}.
The resulting nonsingular, asymptotically flat solutions, often called oscillatons, have fundamental frequencies slightly below the particle mass and can have lifetimes exceeding the Hubble time~\cite{Urena-Lopez:2001zjo, Urena-Lopez:2002ptf, Alcubierre:2003sx, Grandclement:2011wz}.
Gravitational lensing converts this internal oscillation into a direct observable, tying the lensing signal to the dynamics of a self-gravitating real-scalar field~\cite{Boskovic:2018rub, Koutvitsky:2020yfr, Chen:2022kzv}.

The lens map of an extended nonsingular transparent object can contain a radial caustic~\cite{1992grle.book.....S, Dabrowski:1998ac}.
We show that an oscillating boson star can host an \emph{oscillating radial caustic}, with associated radial-critical images formed by light rays passing through the lens interior.
As the stellar configuration oscillates, the radial caustic pulsates in the source plane, so a nearby source crosses it every half period, producing periodic caustic-crossing lensing.
Each crossing can produce an orders-of-magnitude magnification jump~\cite{2002glml.book.....M}.
The resulting observables are phase locked to the lens oscillation: image-pair creation or annihilation near the radial critical curve, changing image morphology, achromatic photometric spikes, and astrometric motion.
This provides a clear and distinctive target for time-domain astronomy.
Other compact lenses lack this full signal: Black holes capture the relevant interior-passing trajectories, opaque baryonic stars block them, and stationary complex-scalar boson stars lack an intrinsically oscillating caustic.
Detecting this lensing phenomenon would provide compelling evidence for an underlying oscillating boson star or a related compact dark-sector object, while null searches would constrain their abundance over wide mass ranges.
We set $c=\hbar=G=1$ throughout the Letter.

\textit{Gravitational lensing---}%
For a spherically symmetric configuration, the metric can be written as
\begin{equation}
    \dif s^2 = -B(t,r) \dif t^2 + A(t,r) \dif r^2 + r^2(\dif\vartheta^2 + \sin^2\vartheta\, \dif\varphi^2) ,
\end{equation}
with two time-dependent metric functions $A(t,r)$ and $B(t,r)$.
A minimally coupled massive real-scalar field~$\phi$ evolving in a curved background is governed by the Einstein-Klein-Gordon (EKG) equations $G_{ab}=\kappa T_{ab}$ and $\nabla^a\nabla_a\phi=m^2 \phi$, which have solutions of the form~\cite{Seidel:1991zh}
\begin{subequations}
    \begin{align}
        A(t,r) &= 1 + \sum_{j=0}^{\jmax} A_{2j}(r) \cos(2j\omega t) , \\
        B(t,r) &= 1 + \sum_{j=0}^{\jmax} B_{2j}(r) \cos(2j\omega t) , \\
        \phi(t,r) &= \sum_{j=1}^{\jmax} \phi_{2j-1}(r) \cos([2j-1]\omega t) ,
    \end{align}
\end{subequations}
where $\omega$ is the fundamental frequency and $\jmax$ is the mode at which the Fourier series is truncated.
Substituting these series into the EKG equations and matching terms of the same frequency yield a set of coupled ordinary differential equations for the radial coefficients.
The boundary conditions are determined by requiring asymptotic flatness and central regularity.
Introducing the dimensionless variables $\tilde{t} \equiv m t$, $\tilde{r} \equiv m r$, $\tilde{\phi} \equiv \sqrt{\kappa}\phi$, and $\tilde{\omega} \equiv \omega/m$, one finds that for every value of $\tilde{\phi}_{1}(0)$ a solution exists~\cite{Seidel:1991zh, Urena-Lopez:2001zjo, Urena-Lopez:2002ptf, Alcubierre:2003sx}.
The mass of oscillating boson stars is given by $M=\tilde{M}/m$ with dimensionless mass $\tilde{M} \equiv \lim_{\tilde{r}\to\infty} \tilde{r}[1-A(\tilde{t},\tilde{r})^{-1}]/2$, which has a maximum value $\tilde{M}_{\max} \approx 0.6$ for stable configurations~\cite{Alcubierre:2003sx}.
Because the Fourier expansion converges rapidly, we retain only the leading oscillatory terms and set $j_{\max}=1$.

\begin{figure}[htbp]
    \centering
    \includegraphics[width=.46\textwidth]{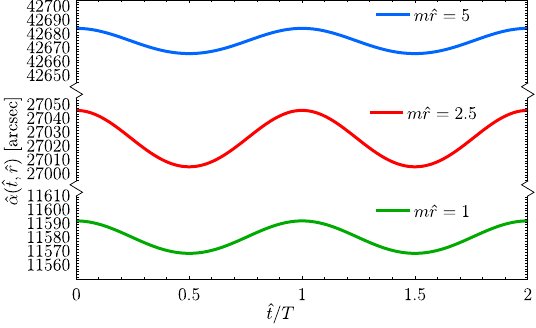}
    \includegraphics[width=.46\textwidth]{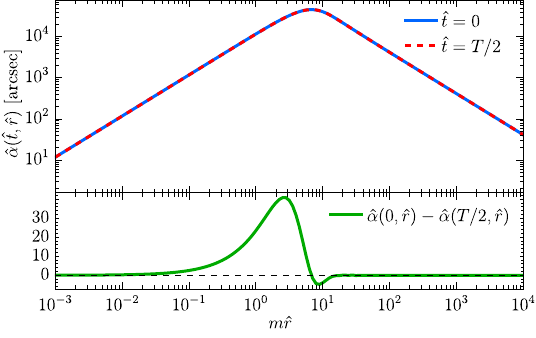}
    \caption{%
        Deflection angle $\alphahat$ as a function of $\{\hat{t}, \hat{r}\}$ for the oscillating boson star with $\tilde{\phi}_{1}(0)=0.133$, which yields $\tilde{M} \approx 0.5$ and $\tilde{\omega} \approx 0.95$.
    }
    \label{fig:alphahat}
\end{figure}

Photon trajectories follow null geodesics.
Restricting to the equatorial plane ($\vartheta=\pi/2$), a geodesic is specified by the time~$\hat{t}$ and radius~$\hat{r}$ of closest approach, which together determine the deflection angle~$\alphahat$.
In Fig.~\ref{fig:alphahat}, we show the corresponding deflection angle as a function of $\{\hat{t}, \hat{r}\}$ for a representative oscillating boson star with $\tilde{\phi}_{1}(0)=0.133$, which yields $\tilde{M} \approx 0.5$ and $\tilde{\omega} \approx 0.95$.
Because the metric oscillates at frequency $2\omega$, the deflection angle has period $T=\pi/\omega$, producing astrometric and photometric image variations.
For a given $\hat{t}$, there is a characteristic radius $\rc$.
For $\hat{r} \gg \rc$ the metric approaches the Schwarzschild form, leading to $\alphahat \simeq {4M}/{\hat{r}}$, whereas for $\hat{r} \ll \rc$ the central regularity gives $\alphahat \propto \hat{r}$.
We find that the numerical deflection angle is accurately described by a validated broken power-law interpolation anchored to the correct interior and exterior asymptotic limits:
\begin{equation}
    \alphahat(\hat{t},\hat{r}) = \frac{4M \hat{r}}{\rc(\hat{t})^2} \left[ 1+ \left( \frac{\hat{r}}{\rc(\hat{t})} \right)^{1/\Delta} \right]^{-2\Delta} ,
    \label{eq:dfl-ang-para}
\end{equation}
where $\Delta \simeq 1/4$ with only mild configuration dependence.
The intrinsic oscillation also makes $\rc$ periodic, and below we adopt Eq.~\eqref{eq:dfl-ang-para}, which captures the essential physics while remaining analytically tractable.

The boson-star radius $R$ is commonly defined as the radius enclosing $99\%$ of the total mass.
The deflection angle grows at small $\hat{r}$ as the enclosed mass increases but decreases for $\hat{r}\gtrsim R$ once the enclosed mass has nearly saturated and the growing impact parameter dominates.
It, therefore, peaks at $\hat{r}$ of the order of $R$.
Since the characteristic radius $\rc$ is defined as the radius at which the deflection angle is maximal, one naturally expects $R/\rc \sim \mathcal{O}(1)$, which we also verify numerically.
The peak deflection angle is $\alphahatpk = 2\sqrt{2}M/\rc \sim 2\sqrt{2}M/R = 2\sqrt{2}\mathcal{C} $, where $\mathcal{C} \equiv M/R$ is the boson-star compactness.
Since $\mathcal{C}$ increases with $\tilde{M}$, the peak deflection increases correspondingly with $\tilde{M}$.
For the most compact stable configuration ($\tilde{M}=\tilde{M}_{\max}$), we find $\alphahatpk \sim 0.4\,\mathrm{rad}\approx 22.9^\circ$.
By contrast, for $\tilde{M} \lesssim 0.1$ one has $\alphahatpk \lesssim 0.01\,\mathrm{rad}\approx 0.57^\circ$.
In this small-amplitude regime, the scale invariance of the EKG equations gives $\tilde{R}\equiv mR\propto \tilde{M}^{-1}$ and, therefore, $\alphahatpk \propto \tilde{M}^{2}$.
See more details in Supplemental Material~\cite{SeeSupplementalMaterial}.

The lensing geometry is described by the lens equation
\begin{equation}
    \beta = \theta - \alpha ,
\end{equation}
where $\beta$ is the angular separation of the source from the optical axis (chosen for convenience along the lens direction), $\theta$ is the observed angular position of an image, and $\alpha$ is the reduced deflection angle, which differs from the geodesic deflection angle $\alphahat$.
At low redshift where redshift effects are negligible, one has~\cite{1987AmJPh..55..428O, Virbhadra:1999nm, Bozza:2008ev, Kudo:2024aak, Hogg:1999ad}
\begin{equation}
    \alpha \equiv \alphahat -\arcsin\left(\frac{\DOL}{\DOS}\big[\sin(\alphahat-\theta)+\sin\theta\big]\right) ,
\end{equation}
and $\theta = \arcsin(\hat{r}/\DOL)$; see the Friedmann-Lema\^{i}tre-Robertson-Walker lens equation in Supplemental Material~\cite{SeeSupplementalMaterial}.
Here, $\DOL$ and $\DOS$ are the observer-lens and observer-source proper distances, respectively.
Introducing the length scale $\mathcal{D} \equiv \sqrt{4M D}$ with $D \equiv \DOL(\DOS-\DOL)/\DOS$, one can define $\nu \equiv \rc/\mathcal{D}$.
Because the characteristic radius $\rc$ oscillates, $\nu(t)$ is periodic in time.
We model its time dependence as $\nu(t) = \bar{\nu}[1-\varepsilon \cos(2\pi t/T)]$, where $\varepsilon$ depends only on $\tilde{M}$, while $\bar{\nu}$ depends on both $\tilde{M}$ and $\tilde{D}\equiv mD$.

\textit{Periodic caustic-crossing lensing---}%
In the caustic-forming regime $\nu<1$, the lens map contains both tangential and radial caustics.
The tangential caustic collapses to the optical-axis point in the source plane and corresponds to the Einstein ring in the image plane.
The boson-star oscillation can, therefore, modulate the Einstein-ring radius and magnification without changing image multiplicity.

\begin{figure}[htbp]
    \centering
    \includegraphics[width=.46\textwidth]{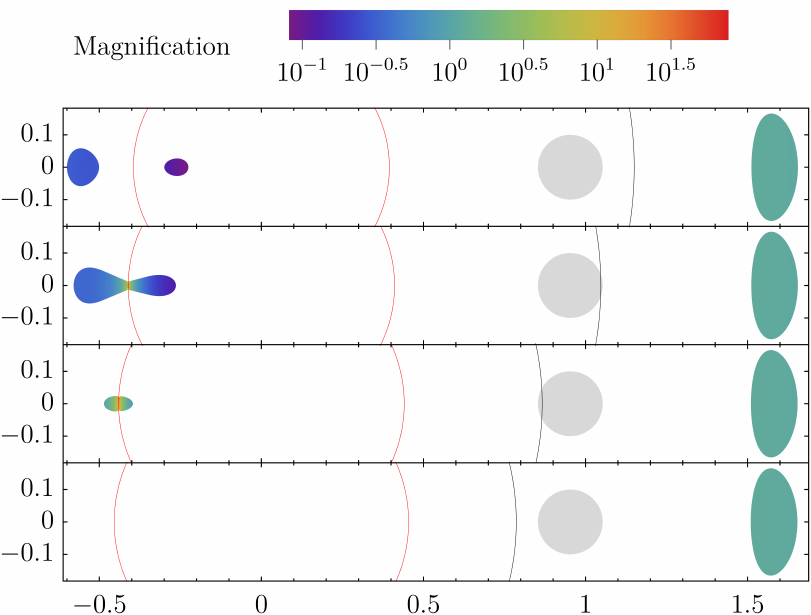}
    \caption{%
        Schematic image morphology for periodic caustic-crossing lensing as the radial caustic contracts over half an oscillation period.
        The gray disk marks the source; the black and red curves indicate the radial caustic and radial critical curve, respectively.
        From top to bottom: in phase~1, the source lies inside the caustic and three images form; in phases~2 and~3, the caustic crosses the source and the image pair near the critical curve approaches and brightens; in phase~4, the source lies outside the caustic and the pair annihilates at the critical curve, leaving one image.
    }
    \label{fig:diagram1}
\end{figure}

The radial caustic, by contrast, is a finite circle in the source plane.
Its angular radius $\betacau(\nu)$ decreases monotonically with $\nu$, so the oscillation of $\nu(t)$ makes the radial caustic periodically pulsate in and out (details in Supplemental Material~\cite{SeeSupplementalMaterial}).
A source lying within the caustic's oscillation range can, therefore, cross it every half period, producing periodic caustic-crossing lensing.
The caustic is absent for $\nu\ge 1$ so that every source position produces a single image.
For $\nu<1$, a source outside the radial caustic yields one image, a source on it produces two images, and a source inside it gives three images.
Moreover, because the deflection profile rises and then falls with the closest-approach radius, the radial critical curve occurs at a smaller closest-approach radius than the tangential critical curve and probes the more strongly oscillating interior potential.
Each caustic crossing induces a sharp magnification spike, the sudden creation or annihilation of a bright image pair at the radial critical curve, and a corresponding change in image morphology.
Figure~\ref{fig:diagram1} presents a schematic illustration of this periodic caustic-crossing lensing produced by an oscillating boson star.
A background source that repeatedly traverses an oscillating caustic, therefore, exhibits a sequence of synchronous astrometric and photometric spikes; detecting such a pattern would constitute strong evidence for an underlying oscillating boson star.

\begin{figure}[htbp]
    \centering
    \includegraphics[width=.46\textwidth]{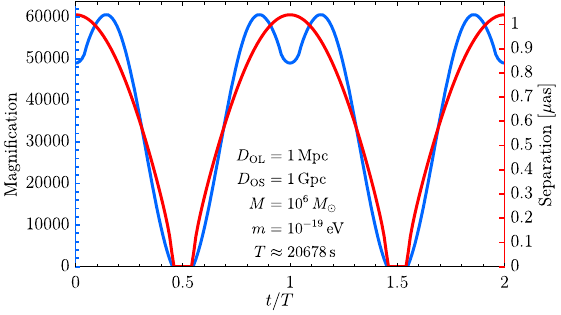}
    \includegraphics[width=.46\textwidth]{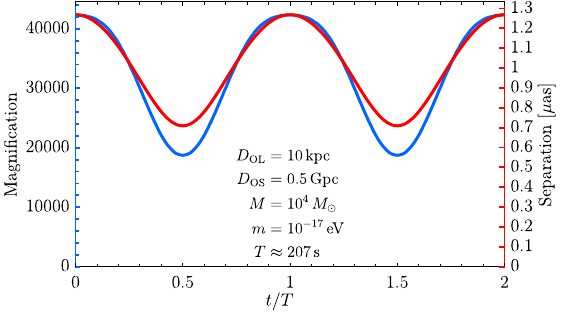}
    \caption{%
        Time evolution of the image pair for two representative lensing configurations.
        The blue curve shows the mean image-pair magnification with finite-source-size effects included ($\Rsrc = 1 R_\odot$, left blue axis), and the red curve shows the pair's maximum radial separation (right red axis).
    }
    \label{fig:img}
\end{figure}

In Fig.~\ref{fig:img} we show the time evolution of the image pair for two representative lensing configurations.
Top panel:
an oscillating boson star with $M=10^6\Msun$ and $m=10^{-19}\eV$ (period $T \approx 20678\second$) at $\DOL=1\Mpc$ lenses a solar-radius source ($\Rsrc = 1 R_\odot$) with uniform surface brightness at $\DOS=1\Gpc$; i.e., both lens and source lie outside the Milky Way.
Here, the source's angular size is smaller than the caustic's angular excursion, so the system passes through all four phases.
The blue curve shows the mean magnification of the image pair with finite-source-size effects included~\cite{1994ApJ...421L..71G,2024SSRv..220...57W}; the red curve shows its maximum radial separation.
The magnification cycles from $\mathcal{O}(10^4)$ down to $0$, while the radial separation oscillates from $\muas$ scales to $0$.
Such extreme magnifications make detection feasible even on cosmological scales, as demonstrated by the $\mathcal{O}(10^{3})$ magnification of an individual star at redshift $1.5$ by a galaxy-cluster lens~\cite{Kelly:2017fps}.
Bottom panel:
lens ($M=10^4\Msun$, $m=10^{-17}\eV$, $T \approx 207\second$) at $\DOL=10\kpc$, with the same source at $\DOS=0.5\Gpc$ (lens inside, source outside the Milky Way).
Here the source's angular size exceeds the caustic's excursion, and the source position is chosen so that the system samples only phases~2 and~3.
Because phase~4 does not occur, the minimum magnification and separation remain nonzero.
For the two examples, ignoring redshift effects is adequate for the illustrations and order-of-magnitude estimates; both have the same dimensionless mass $\tilde{M}\approx 7.5\times10^{-4}$ and compactness $\mathcal{C}\approx 5.7\times10^{-8}$.
They are, therefore, dilute, small-amplitude examples far from the maximum-mass branch and are not intended to represent the full oscillating-boson-star population.
They instead illustrate two observational geometries in the stable dilute regime, while the event-number estimate below scans the broader parameter space.
More generally, the variation strength depends on the boson-star compactness.
More compact boson stars modulate $\rc$ more strongly, which increases the caustic excursion and correspondingly enhances the periodic photometric and astrometric changes.

\begin{figure}[htbp]
    \centering
    \includegraphics[width=.46\textwidth]{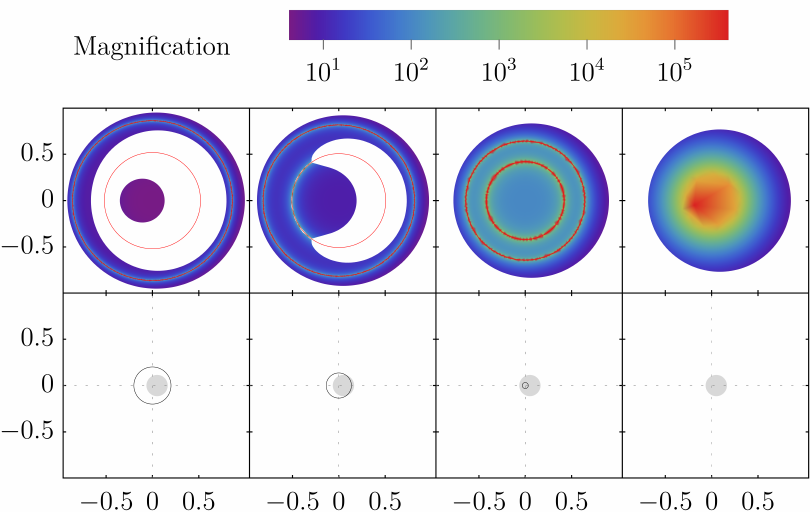}
    \caption{%
        Schematic image morphology for type-II periodic caustic-crossing lensing as the radial caustic contracts over half an oscillation period.
        The gray disk marks the source; the black curve is the radial caustic, the inner red curve is the radial critical curve, and the outer red curve is the Einstein ring.
        From left to right: in phase~1, the source lies inside the caustic, producing a bright Einstein ring and central image; in phase~2, the caustic crosses the source and the image pair near the critical curve approaches and brightens; in phase~3, the radial caustic lies inside the source and the morphology contains an outer Einstein ring and an inner radial-critical ring; in phase~4, the caustic vanishes ($\nu>1$), the bright rings disappear, and a single image with a bright central region remains.
    }
    \label{fig:diagram2}
\end{figure}

For an extended source, a second type of periodic lensing arises when the source covers the lens center, as illustrated in Fig.~\ref{fig:diagram2}.
We refer to this as type-II periodic caustic-crossing lensing, in contrast to type~I in Fig.~\ref{fig:diagram1} where the source does not cover the lens center.
In the type-II case, whenever a caustic is present, an Einstein ring forms, and the resulting image morphology differs markedly from type~I.

\textit{Event number---}%
For periodic caustic-crossing lensing to occur, two conditions must be satisfied:
(i) an oscillating radial caustic must exist, i.e.~$\nu_{\min}<1$, which implies
\begin{subequations}
    \begin{align}
        \DOL &> \rcS , \\
        \DOS &> \DOL^2/(\DOL-\rcS) ,
    \end{align}
\end{subequations}
where $\rcS \equiv r_{\mathrm{c},\min}^2/(4M)\sim R/(4\mathcal{C})$---this makes explicit that caustic formation is favored for more compact boson-star configurations;
(ii) the source must intersect the caustic at some phase, which constrains its angular position to $\beta_{\min}<\beta_{\src}<\beta_{\max}$, with
\begin{subequations}
    \begin{equation}
        \mathrm{type~I:} \left\{
            \begin{aligned}
                \beta_{\min} &= \max\left[ \frac{R_{\src}}{\DOS},\, \betacau(\nu_{\max})-\frac{R_{\src}}{\DOS} \right], \\
                \beta_{\max} &= \betacau(\nu_{\min})+\frac{R_{\src}}{\DOS} ,
            \end{aligned}
        \right.
    \end{equation}
    \begin{equation}
        \mathrm{type~II:} \left\{
            \begin{aligned}
                \beta_{\min} &= \max\left[ 0,\, \betacau(\nu_{\max})-\frac{R_{\src}}{\DOS} \right], \\
                \beta_{\max} &= \frac{R_{\src}}{\DOS} ,
            \end{aligned}
        \right.
    \end{equation}
\end{subequations}
where $R_{\src}$ is the source radius.

For survey depth $Z$ and number densities $n_a$ and $n_b$ of oscillating boson stars and sources, the expected number of periodic caustic-crossing lensing events is
\begin{equation}
    \begin{aligned}
        N_{\evt}
 &= \int_0^Z \dif\DOL \bigg\{ n_a \,\Theta(\DOL-\rcS) \,4\pi\DOL^2 \\
 &\quad \times \int_{\DOL}^{Z} \dif\DOS \Big\{ n_b \,\Theta(\DOS-\frac{\DOL^2}{\DOL-\rcS}) \\
 &\quad \times 2\pi\DOS^2 \; \max\left[0,\, \cos\beta_{\min} - \cos\beta_{\max} \right] \Big\} \bigg\} \\
 &\equiv \left(\frac{n_a}{\kpc^{-3}}\right) \left(\frac{n_b}{\kpc^{-3}}\right) \; \mathcal{N}(m,M,Z,\Rsrc) ,
    \end{aligned}\label{eq:Nevt}
\end{equation}
where $\Theta(x)$ is the Heaviside step function.
For type~I, the inequality $\beta_{\min}<\beta_{\max}$ is automatically satisfied.
For type~II, this inequality imposes $\betacau(\nu_{\max})<2{R_{\src}}/{\DOS}$, which sets an additional upper bound on $\DOS$.
In practical surveys $R_{\src}\ll Z$, so the effective type-II upper limit on $\DOS$ is much smaller than $Z$; the type-II yield is, therefore, far below type~I, and we focus on type~I below.

\begin{figure}[t]
    \centering
    \includegraphics[width=.46\textwidth]{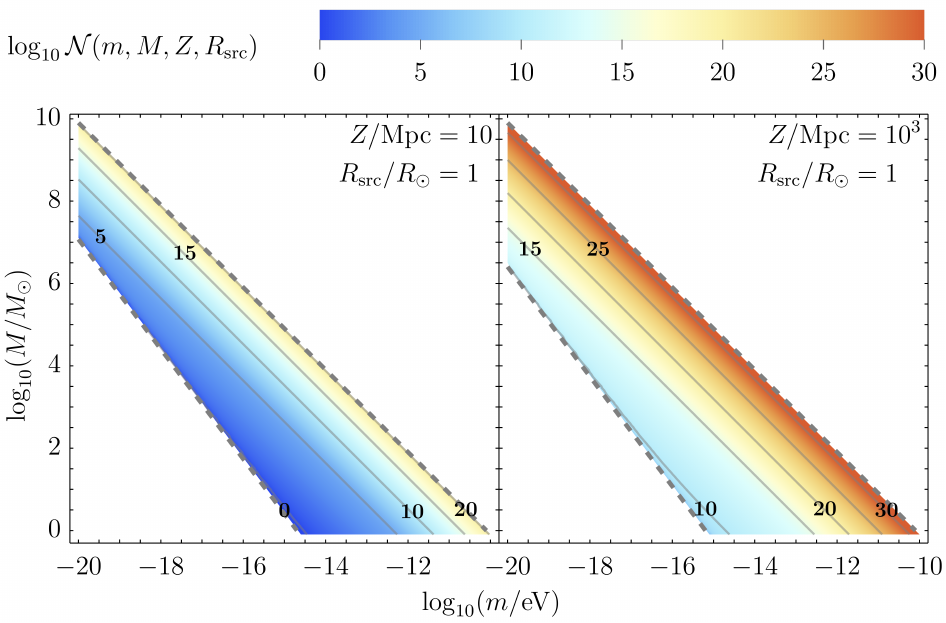}
    \caption{%
        Reduced event number $\mathcal{N}$ for type-I periodic caustic-crossing lensing as a function of model parameters.
    }
    \label{fig:lgN}
\end{figure}

Figure~\ref{fig:lgN} displays the reduced event number $\mathcal{N}$ for a range of parameter choices.
The upper boundary corresponds to the stability limit $M_{\max} = \tilde{M}_{\max}/m \approx 8\times10^9 \Msun \left({10^{-20}\eV}/{m}\right)$, while the requirement of a radial caustic ($\nu<1$) gives the lower bound $M_{\min} \approx 1.6 \times 10^7 \Msun \left({Z/4}\,/{\Mpc}\right)^{-1/3} \left({10^{-20}\eV}/{m}\right)^{4/3}$.
These bounds are indicated by the dashed lines.
The parametric dependence of the reduced event number can be written approximately as $\mathcal{N}(m,M,Z,\Rsrc) \sim Z^6 F_1(\tilde{M}) F_2(\Rsrc)$, where $F_2(\Rsrc) \sim \mathcal{O}(1)$ for realistic stellar sizes.
This scaling is intuitive: The volume integral $\int^Z\DOL^2\dif\DOL \int^Z\DOS^2\dif\DOS$ contributes the $Z^6$ factor; the lensing configuration, which is scale invariant under $m$, supplies the $F_1(\tilde{M})$ dependence; and the practical regime $\Rsrc \ll D$ makes $F_2(\Rsrc)$ order unity.
For a survey extending to $Z=1\Gpc$, the reduced event number $\mathcal{N}$ can be as large as $\mathcal{O}(10^{30})$, which implies a correspondingly large event number $N_{\evt}$ once realistic values of $n_a$ and $n_b$ are applied.
This estimate counts all periodic caustic-crossing events irrespective of image brightness; consequently, a substantial fraction will be too faint for current facilities.

\begin{figure}[t]
    \centering
    \includegraphics[width=.46\textwidth]{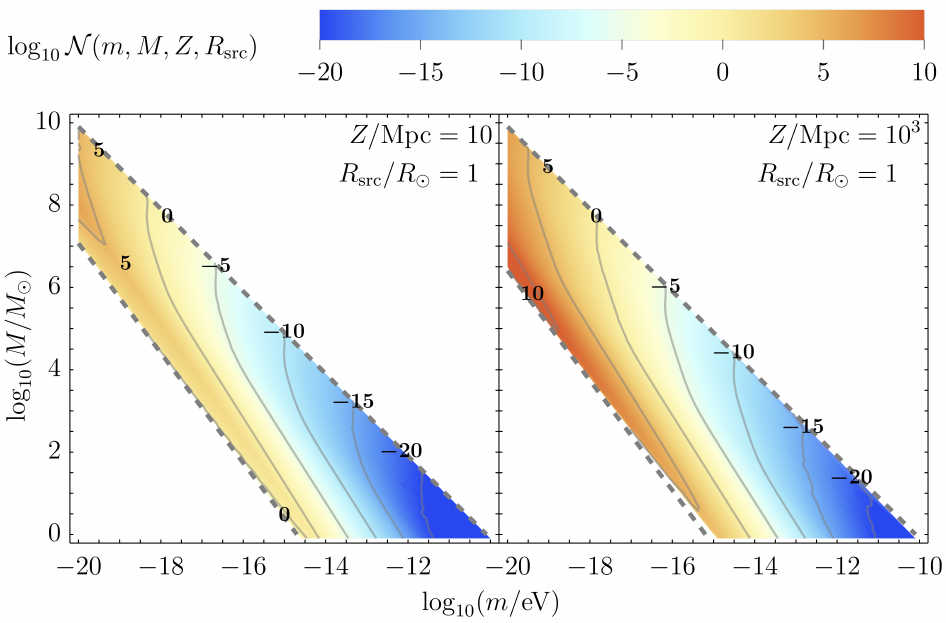}
    \caption{%
        Reduced event number $\mathcal{N}$ after applying a fiducial detection threshold (see the text).
    }
    \label{fig:lgN_criteria}
\end{figure}

To estimate the number of detectable events for current facilities, consider a solar-radius source with absolute magnitude $M_{\abs}=0$.
Its apparent magnitude is $m_{\app}=M_{\abs}-2.5\log_{10} [\left({10\pc}/{\DOS}\right)^{2}\mu ]$, where $\mu$ is the lensing magnification averaged over the finite size of the source.
Imposing a detection limit $m_{\app}=30$ requires $\left({10\Mpc}/{\DOS}\right)^{2}\,\mu > 1$.
Including this criterion produces the reduced event number shown in Fig.~\ref{fig:lgN_criteria}.

As a concrete illustration, take fuzzy dark matter to be the real-scalar field analyzed above~\cite{Ferreira:2020fam}.
If a fraction $f_{\mathrm{BS}}$ of dark matter resides in boson stars of mass $M$, the number density of boson stars is $n_a = f_{\mathrm{BS}}\rho_{\DM}/M$, where $\rho_\DM$ is the density of dark matter appropriate to the survey volume.
Adopting a representative mean stellar number density $n_b \sim 1\kpc^{-3}$~\cite{Madau:2014bja} and taking $m=10^{-18}\eV$, $M=10^4\Msun$, and $f_{\mathrm{BS}}=10^{-3}$, the expected number of detectable periodic caustic-crossing events in a survey extending to $Z=1\Gpc$ is $N_{\evt} \sim \mathcal{O}(10^3)$.

All of the foregoing analysis extends straightforwardly to a massive real scalar with self-interactions~\cite{Urena-Lopez:2012udq}.
In that setting, axionlike particles provide an especially well-motivated example~\cite{Svrcek:2006yi, Arvanitaki:2009fg}.
Axion stars composed of such particles produce analogous periodic lensing signatures.
For a dilute axion star, one finds that $m=10^{-18}\eV$ can yield $M=10^4\Msun$ with a boson-star fraction $f_{\mathrm{BS}} \simeq 0.4$~\cite{Chang:2024fol, Braaten:2019knj}, which, in turn, implies an expected number of detectable periodic caustic-crossing events of the order of $N_{\evt} \sim \mathcal{O}(10^6)$ for a survey depth of $Z=1\Gpc$.

\textit{Discussion---}%
We have shown that oscillating real-scalar boson stars generically host an oscillating radial caustic.
A source near this caustic crosses it every half period, producing periodic caustic-crossing lensing with image-pair creation or annihilation, changing image morphology, achromatic photometric spikes, and astrometric motion.
This identifies an intrinsically time-dependent caustic structure and a caustic-driven lensing signal phase locked to the coherent oscillation of the lens itself.
We also derived an event-number estimate and, after imposing a fiducial magnitude cut, found that a measurable subset of such events could be detectable with current facilities.
Null searches would constrain the abundance of these compact dark objects across wide mass ranges.
These predictions depend only on the dynamics of real-scalar condensates and, therefore, provide a clean lensing test of self-gravitating quantum fields in curved spacetime.
The framework naturally extends to self-interacting real scalars, including axionlike particles, and to ultralight vector bosons~\cite{Jain:2021pnk, Zhang:2021xxa}, for which the core predictions of an oscillating caustic and periodic observables remain intact.
More broadly, exotic bosonic sectors with oscillatory dynamics suggest that time-dependent lensing may be a useful probe beyond the minimal oscillaton model~\cite{Jaramillo:2023twi, Bernal:2026new}.

The periodic caustic-crossing lensing uncovered here is a distinctive target for time-domain astronomy.
Its periodicity is intrinsic to the lens map, not produced by transverse source-lens motion.
This phase-locked caustic motion provides a discriminant against ordinary lensing variability and motivates searches combining photometric and astrometric time series.
Detecting it requires a suitable combination of photometric depth and precision, high-cadence sampling, a long enough time baseline, and (ideally) sufficient angular resolution.
In the near term, joint analyses that combine Gaia-grade astrometry with high-cadence photometric surveys can already probe periods in the $T \sim 10^2-10^5\,\mathrm{s}$ window and place meaningful null constraints on the number density of oscillating boson stars.
Looking ahead, improved microarcsecond astrometry and deeper, faster time-domain coverage will extend the sensitivity toward lower masses and shorter periods.
While we focus on optical gravitational lensing in this Letter, gravitational waves can also be lensed by oscillating boson stars.
As nonstatic lenses with caustics, oscillating boson stars can imprint distinctive signatures on gravitational-wave signals~\cite{Yang:2024rwc, Ezquiaga:2025gkd}, which we leave for future work.

\textit{Acknowledgments---}%
X.-Y.~Y. is supported in part by the KIAS Individual Grant No.~QP090702.
T.~C. is supported by the National Natural Science Foundation of China Grant No.~12547155.
R.-G.~C. is supported in part by the National Natural Science Foundation of China Grant No.~12588101.

\bibliography{citeLib}

\onecolumngrid
\begin{center}
    \vspace{2em}
    {\large\bfseries --- Supplemental Material --- \par}
    \vspace{1em}
\end{center}
\twocolumngrid

\setcounter{equation}{0}
\renewcommand{\theequation}{S\arabic{equation}}

\setcounter{figure}{0}
\renewcommand{\thefigure}{S\arabic{figure}}

\setcounter{table}{0}
\renewcommand{\thetable}{S\arabic{table}}

\begin{figure}[htbp]
    \centering
    \includegraphics[width=.46\textwidth]{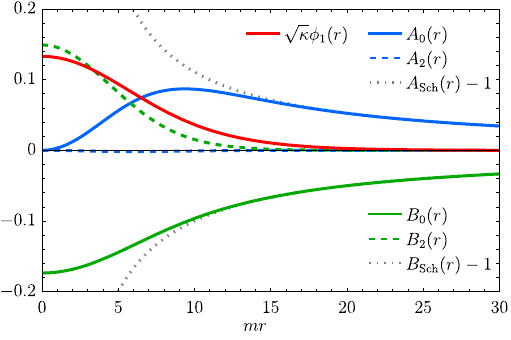}
    \caption{%
    Representative oscillating boson-star solution with $\tilde{\phi}_{1}(0)=0.133$, giving $\tilde{M}\approx 0.4995$ and $\tilde{\omega}\approx 0.9528$.
    Here $A_\mathrm{Sch}$ and $B_\mathrm{Sch}$ denote the metric functions of the Schwarzschild metric.
    }
    \label{fig:sol}
\end{figure}

\textbf{Oscillating boson-star solution.}
For a spherically symmetric configuration, we write
\begin{equation}
    \dif s^2 = -B(t,r)\dif t^2 + A(t,r)\dif r^2 + r^2(\dif\vartheta^2+\sin^2\vartheta\,\dif\varphi^2) .
\end{equation}
A minimally coupled massive real scalar field~$\phi$ evolving in this background is governed by the Einstein-Klein-Gordon equations $G_{ab}=\kappa T_{ab}$ and $\nabla^a\nabla_a\phi=m^2\phi$.
In terms of the dimensionless variables $\tilde{t}\equiv mt$, $\tilde{r}\equiv mr$, and $\tilde{\phi}\equiv \sqrt{\kappa}\phi$, these equations read
\begin{subequations}\label{eq:EKG_dimless}
\begin{align}
    \frac{A'}{A} &=
    \frac{1-A}{\tilde{r}}
    + \frac{\tilde{r}}{2}
    \left(\frac{A}{B}\dot{\tilde{\phi}}^2+\tilde{\phi}'^2+A\tilde{\phi}^2\right) ,
    \\
    \frac{B'}{B} &=
    \frac{A-1}{\tilde{r}}
    + \frac{\tilde{r}}{2}
    \left(\frac{A}{B}\dot{\tilde{\phi}}^2+\tilde{\phi}'^2-A\tilde{\phi}^2\right) ,
    \\
    \ddot{\tilde{\phi}} &=
    \left(\frac{\dot{B}}{2B}-\frac{\dot{A}}{2A}\right)\dot{\tilde{\phi}}
    + \frac{B}{A}\tilde{\phi}''
    \nonumber\\
    &\quad
    + \left(\frac{B'}{2B}-\frac{A'}{2A}+\frac{2}{\tilde{r}}\right)
    \frac{B}{A}\tilde{\phi}'
    - B\tilde{\phi} ,
\end{align}
\end{subequations}
where overdots and primes denote derivatives with respect to $\tilde{t}$ and $\tilde{r}$, respectively.
We use the Fourier expansion
\begin{subequations}\label{eq:sol_form_dimless}
\begin{align}
    A(\tilde{t},\tilde{r}) &=
    1 + \sum_{j=0}^{\jmax} A_{2j}(\tilde{r})\cos(2j\tilde{\omega}\tilde{t}) ,
    \\
    B(\tilde{t},\tilde{r}) &=
    1 + \sum_{j=0}^{\jmax} B_{2j}(\tilde{r})\cos(2j\tilde{\omega}\tilde{t}) ,
    \\
    \tilde{\phi}(\tilde{t},\tilde{r}) &=
    \sum_{j=1}^{\jmax} \tilde{\phi}_{2j-1}(\tilde{r})
    \cos([2j-1]\tilde{\omega}\tilde{t}) ,
\end{align}
\end{subequations}
with $\tilde{\omega}\equiv \omega/m$.
Substituting Eqs.~\eqref{eq:sol_form_dimless} into Eqs.~\eqref{eq:EKG_dimless} and matching terms of the same frequency yields a set of coupled ordinary differential equations for the radial coefficients $A_{2j}(\tilde{r})$, $B_{2j}(\tilde{r})$, and $\tilde{\phi}_{2j-1}(\tilde{r})$.
We retain only the leading oscillatory terms and set $\jmax=1$, because the Fourier expansion converges rapidly.
The boundary conditions are determined by asymptotic flatness and central regularity.
Asymptotic flatness requires $\tilde{\phi}|_{\tilde{r}\to\infty}\to0$ and $\{A,B\}|_{\tilde{r}\to\infty}\to1$.
Central regularity implies $\partial_{\tilde{r}}\tilde{\phi}|_{\tilde{r}=0}=0$ and $A|_{\tilde{r}=0}=1$.
A representative solution with central amplitude $\tilde{\phi}_{1}(0)=0.133$ is shown in Fig.~\ref{fig:sol}; it has dimensionless mass $\tilde{M}\approx 0.4995$ and frequency $\tilde{\omega}\approx 0.9528$.
For this solution, the central values of the metric coefficients are $B_0(0)\approx -0.1735$ and $B_2(0)\approx 0.1491$.
The coefficient $A_0$ reaches a maximum value $A_0\approx 0.08696$ at $\tilde{r}\approx 9.364$, while $A_2$ reaches a minimum value $A_2\approx -0.001697$ at $\tilde{r}\approx 5.961$.

In the small-amplitude regime $\tilde{\phi}_{1}(0) \ll 1$, the equations exhibit an approximate scale invariance. Specifically, for any $\ell>0$,
\begin{equation}
    \begin{aligned}
        &\{ \phantom{\ell^{-1}} \tilde{r},\ \phantom{\ell^2} \tilde{\phi}_{1},\ \phantom{\ell^2} B_0,\ \phantom{\ell^2} B_2,\ \phantom{\ell^2} A_0,\ \phantom{\ell^4} A_2 \} \\
        \rightarrow &\{ \ell^{-1} \tilde{r},\ \ell^2 \tilde{\phi}_{1},\ \ell^2 B_0,\ \ell^2 B_2,\ \ell^2 A_0,\ \ell^4 A_2 \} ,
    \end{aligned}
\end{equation}
which underlies the familiar mass-radius scaling of oscillating boson-star solutions, $\tilde{R}\propto \tilde{M}^{-1}$.

\begin{table*}[t]
    \caption{%
    Extrema of the deflection-angle curves shown in Fig.~1 of the main text for the representative oscillating boson star in Supplemental Fig.~\ref{fig:sol}.
    Values are given in arcseconds, with degrees shown in parentheses.
    }
    \label{tab:alphahat-extrema}
    \begin{ruledtabular}
    \begin{tabular}{@{}lcccc@{}}
        Curve & Max. value & Max. position & Min. value & Min. position \\
        $\alphahat(\hat{t},\hat{r})$ with $m\hat{r}=5$ & $42684.9\,\mathrm{arcsec}\;(11.86^\circ)$ & $\hat{t}/T=0$ & $42666.6\,\mathrm{arcsec}\;(11.85^\circ)$ & $\hat{t}/T=1/2$ \\
        $\alphahat(\hat{t},\hat{r})$ with $m\hat{r}=2.5$ & $27046.6\,\mathrm{arcsec}\;(7.513^\circ)$ & $\hat{t}/T=0$ & $27005.8\,\mathrm{arcsec}\;(7.502^\circ)$ & $\hat{t}/T=1/2$ \\
        $\alphahat(\hat{t},\hat{r})$ with $m\hat{r}=1$ & $11592.7\,\mathrm{arcsec}\;(3.220^\circ)$ & $\hat{t}/T=0$ & $11569.3\,\mathrm{arcsec}\;(3.214^\circ)$ & $\hat{t}/T=1/2$ \\
        $\alphahat(0,\hat{r})$ & $45403.4\,\mathrm{arcsec}\;(12.61^\circ)$ & $m\hat{r}\approx 7.079$ & $--$ & $--$ \\
        $\alphahat(0,\hat{r})-\alphahat(T/2,\hat{r})$ & $40.8039\,\mathrm{arcsec}\;(0.01133^\circ)$ & $m\hat{r}\approx 2.512$ & $--$ & $--$ \\
    \end{tabular}
    \end{ruledtabular}
\end{table*}

\textbf{Deflection angle.}
Photon trajectories follow null geodesics.
Restricting to the equatorial plane ($\vartheta=\pi/2$), the geodesic equations read
\begin{subequations}
\begin{align}
    0 &= t'' + \frac{\partial_r B}{B} t' r' + \frac{\partial_t A}{2B} r'^2 + \frac{\partial_t B}{2B} t'^2 ,
    \\
    0 &= r'' + \frac{\partial_t A}{A} t' r' + \frac{\partial_r B}{2A} t'^2 + \frac{\partial_r A}{2A} r'^2 - \frac{r}{A}\varphi'^2 ,
    \\
    0 &= \varphi'' + \frac{2}{r}r'\varphi' .
\end{align}
\end{subequations}
Here, in the geodesic equations, primes denote derivatives with respect to an affine parameter $\varsigma$.
We specify a given null geodesic by the time $\hat{t}$ and radius $\hat{r}$ of closest approach and impose the initial data
\begin{subequations}
\begin{align}
    & \left\{ t(\varsigma), r(\varsigma), \vartheta(\varsigma), \varphi(\varsigma) \right\}\big|_{\varsigma=0} = \{ \hat{t}, \hat{r}, \pi/2, 0 \} ,
    \\
    & \left\{ \frac{\dif t}{\dif\varsigma}, \frac{\dif r}{\dif\varsigma}, \frac{\dif \vartheta}{\dif\varsigma}, \frac{\dif \varphi}{\dif\varsigma} \right\}\bigg|_{\varsigma=0} = \{ 1, 0, 0, \frac{\sqrt{B(\hat{t},\hat{r})}}{\hat{r}} \} ,
\end{align}
\end{subequations}
which fix the trajectory and its impact parameter.
The deflection angle is then
\begin{equation}
    \alphahat(\hat{t},\hat{r}) \equiv \varphi(+\infty) - \varphi(-\infty) - \pi .
\end{equation}

\begin{figure}[htbp]
    \centering
    \includegraphics[width=.46\textwidth]{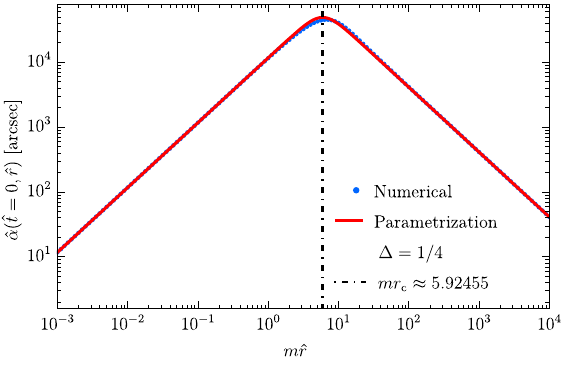}
    \caption{%
    Numerical results and broken power-law parameterization of the deflection angle for the representative oscillating boson star in Fig.~\ref{fig:sol}.
    }
    \label{fig:alphahat_num-para}
\end{figure}

In Fig.~\ref{fig:alphahat_num-para}, we show the numerical results of the deflection angle for the representative oscillating boson star in Fig.~\ref{fig:sol}, together with the broken power-law parameterization in the main text.
The parameterization reproduces the numerical results to high accuracy.
The corresponding extrema of the curves shown in Fig.~1 of the main text are collected in Table~\ref{tab:alphahat-extrema}.

\textbf{Lens equation.}
We formulate the lens equation in a spatially flat FLRW background, which is sufficient for the lensing configurations considered in this work.
The metric is
\begin{equation}
	\dif s^2
	=
	-\dif t^2 + a^2(t) \left[
	\dif\chi^2+\chi^2(\dif\vartheta^2+\sin^2\vartheta\,\dif\varphi^2)
	\right] ,
\end{equation}
where $\chi$ is the comoving radial coordinate.
For a spatially flat FLRW universe, the constant-time comoving spatial slices are Euclidean, and the spatial projections of undeflected null geodesics are straight lines in comoving coordinates.

\begin{figure}[htbp]
	\centering
	\includegraphics[width=.46\textwidth]{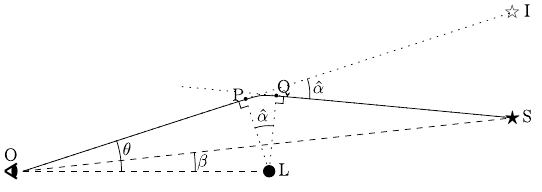}
	\caption{%
		Schematic lensing geometry in comoving coordinates.
	}
	\label{fig:diagram_lens}
\end{figure}

Let the lens and source redshifts be $z_{\mathrm{L}}$ and $z_{\mathrm{S}}$, with corresponding comoving distances
\begin{equation}
	\chi_{\mathrm{L}}=\chi(z_{\mathrm{L}}),\qquad
	\chi_{\mathrm{S}}=\chi(z_{\mathrm{S}}) .
\end{equation}
The angular-diameter distances from the observer to the lens, from the observer to the source, and from the lens to the source are given by~\cite{Hogg:1999ad}
\begin{equation}\label{eq:defangdis}
	d_{\mathrm{OL}}=\frac{\chi_{\mathrm{L}}}{1+z_{\mathrm{L}}},\quad
	d_{\mathrm{OS}}=\frac{\chi_{\mathrm{S}}}{1+z_{\mathrm{S}}},\quad
	d_{\mathrm{LS}}=\frac{\chi_{\mathrm{S}}-\chi_{\mathrm{L}}}{1+z_{\mathrm{S}}} .
\end{equation}

Working in a comoving Cartesian frame with origin at the observer $O$, we write
\begin{subequations}
	\begin{align}
		\overrightarrow{\mathrm{OL}} &= \Big(\chi_{\mathrm{L}},0\Big) \,,\\
		\overrightarrow{\mathrm{OS}} &= \chi_{\mathrm{S}}\Big(\cos\beta,\sin\beta\Big) \,,\\
		\overrightarrow{\mathrm{LQ}} &= |\overrightarrow{\mathrm{LQ}}|
		\Big( \cos(\pi-\angle\mathrm{OLQ}), \sin(\pi-\angle\mathrm{OLQ}) \Big) \,,\\
		\overrightarrow{\mathrm{SQ}} &=
		\overrightarrow{\mathrm{OL}} + \overrightarrow{\mathrm{LQ}} -\overrightarrow{\mathrm{OS}} \,.
	\end{align}
\end{subequations}
Here $\beta$ is the angular source position relative to the optical axis, and $\theta$ is the observed image angle.
The angle at the lens is
\begin{equation}
	\angle\mathrm{OLQ}=\frac{\pi}{2}-\theta+\alphahat .
\end{equation}
The equality of the incoming and outgoing impact parameters gives
\begin{equation}
	|\overrightarrow{\mathrm{LQ}}|
	=
	|\overrightarrow{\mathrm{LP}}|
	=
	\chi_{\mathrm{L}}\sin\theta .
\end{equation}
This follows from the time-reversal symmetry of the local null geodesic and the standard asymptotic approximation outside the lens region.
The physical impact radius entering the local deflection angle is
\begin{equation}
	\hat r
	=
	a_{\mathrm{L}}|\overrightarrow{\mathrm{LP}}|
	=
	\frac{\chi_{\mathrm{L}}}{1+z_{\mathrm{L}}}\sin\theta
	=
	d_{\mathrm{OL}}\sin\theta ,
\end{equation}
so that
\begin{equation}
	\alphahat=\alphahat(\hat t,d_{\mathrm{OL}}\sin\theta).
\end{equation}
Here $\hat t$ denotes the local lens-frame time at which the photon reaches closest approach. The deflection angle is a local quantity determined by the spacetime geometry near the lens. For a lens at redshift $z_{\mathrm{L}}$, time intervals measured by the observer are redshifted according to $\Delta t_{\rm obs}=(1+z_{\mathrm{L}})\Delta\hat t$.

Since $\overrightarrow{\mathrm{LQ}}$ is orthogonal to $\overrightarrow{\mathrm{SQ}}$, we have
\begin{equation}
	\begin{aligned}
		0 =
		\overrightarrow{\mathrm{LQ}}\cdot\overrightarrow{\mathrm{SQ}} =
		\chi_{\mathrm{L}}\sin\theta
		\Big\{ &
		\chi_{\mathrm{L}}
		\big[
		\sin(\alphahat-\theta)+\sin\theta
		\big] \\
		&-
		\chi_{\mathrm{S}}
		\sin(\alphahat+\beta-\theta)
		\Big\},
	\end{aligned}
\end{equation}
which yields
\begin{equation}
	\sin(\alphahat+\beta-\theta)
	=
	\frac{\chi_{\mathrm{L}}}{\chi_{\mathrm{S}}}
	\big[
	\sin(\alphahat-\theta)+\sin\theta
	\big] .
\end{equation}
Therefore, the lens equation in a spatially flat FLRW universe is
\begin{equation}
	\beta
	=
	\theta-\alphahat
	+
	\arcsin
	\left(
	\frac{\chi_{\mathrm{L}}}{\chi_{\mathrm{S}}}
	\big[
	\sin(\alphahat-\theta)+\sin\theta
	\big]
	\right) ,
	\label{eq:flat_FLRW_finite_angle_chi}
\end{equation}
which can be written in terms of angular-diameter distances as
\begin{equation}
	\beta
	=
	\theta-\alphahat
	+
	\arcsin
	\left(
	\big[
	1-\frac{d_{\mathrm{LS}}}{d_{\mathrm{OS}}}
	\big]
	\big[
	\sin(\alphahat-\theta)+\sin\theta
	\big]
	\right) .
	\label{eq:flat_FLRW_finite_angle_d}
\end{equation}

Taking the small-angle limit of Eq.~\eqref{eq:flat_FLRW_finite_angle_d} gives
\begin{equation}
	\beta = \theta - 	\frac{d_{\mathrm{LS}}}{d_{\mathrm{OS}}}\alphahat,
	\label{eq:small_angle_cosmological_lens}
\end{equation}
with $\theta \simeq \hat r / d_{\mathrm{OL}}$.
Thus we recover the standard small-angle cosmological lens equation.

In the low-redshift limit, one has
\begin{subequations}
    \begin{align}
    &d_{\mathrm{OL}}\simeq \chi_{\mathrm{L}} \simeq \DOL,\\
	&d_{\mathrm{OS}}\simeq \chi_{\mathrm{S}} \simeq \DOS,\\
	&d_{\mathrm{LS}}\simeq \chi_{\mathrm{S}}-\chi_{\mathrm{L}} \simeq \DOS-\DOL \equiv \DLS.
    \end{align}
\end{subequations}
The lens equation then reduces to
\begin{equation}
	\beta
	=
	\theta-\alphahat
	+
	\arcsin
	\left(
	\frac{\DOL}{\DOS}
	\big[
	\sin(\alphahat-\theta)+\sin\theta
	\big]
	\right) .
    \label{eq:large_angle_noncosmological_lens}
\end{equation}
This is equivalent to Bozza's improved form of the Ohanian lens equation, which has been shown to provide high accuracy among commonly used lens equations~\cite{Bozza:2008ev,Kudo:2024aak}.

Taking both the low-redshift and small-angle limits gives
\begin{equation}
	\beta = \theta -  \frac{\DLS}{\DOS} \alphahat ,
\end{equation}
which is the usual small-angle lens equation in the asymptotically flat geometry.
In this same limit, the boson-star lens equation can be cast in the compact dimensionless form
\begin{equation}\label{eq:lens_dimless}
	y = x - \nu^{-2}x \left[ 1+ (x/\nu)^{4} \right]^{-1/2},
\end{equation}
with $y \equiv \beta\DOL/\mathcal{D}$ and $x \equiv \theta\DOL/\mathcal{D}$.
Comparing this approximation with the lens equation~\eqref{eq:large_angle_noncosmological_lens}, we find percent-level agreement.
This agreement reflects that the only small-angle substitutions in the lens equation are $\sin\theta \simeq \theta$ and $\sin(\alphahat-\theta) \simeq \alphahat-\theta$.
Even for the most extreme stable configuration ($\tilde{M}=\tilde{M}_{\max}$, $\alphahatpk \sim 0.4\,\mathrm{rad}$), the maximal fractional error introduced by $\sin x \simeq x$ is bounded by $({x}/{\sin x}-1) |_{x=0.4} \approx 2.7\%$.
While less accurate in principle, the small-angle approximation is useful because it yields a compact, scale-free lens equation, making the underlying parameter dependence and physical picture especially transparent.

In this work, we use the low-redshift lens equation~\eqref{eq:large_angle_noncosmological_lens} for the numerical results, which is adequate for the order-of-magnitude estimates of the examples considered.
The FLRW derivation above shows how the lens equation should be generalized when cosmological distance effects are retained.

\begin{figure}[htbp]
    \centering
    \includegraphics[width=.46\textwidth]{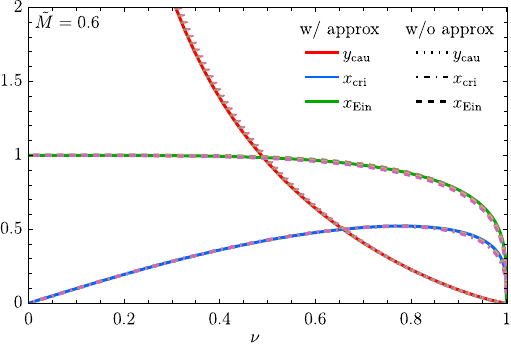}
    \caption{%
    Radial caustic $\ycau$ (source plane), radial and tangential critical curves $\xcri$ and $\xEin$ (image plane) as functions of $\nu$.
    \emph{Solid curves}: small-deflection approximation, in which $\ycau$, $\xcri$ and $\xEin$ depend only on $\nu$.
    \emph{Other curves}: full large-deflection treatment, in which they depend on $\{ \xi,\tilde{M},\nu \}$ with $\xi\equiv \DOL/\DOS$.
    We plot $\xi=\{0.1,0.2,\ldots,0.9\}$.
    The two treatments agree at the percent level across the parameter range shown.
    }
    \label{fig:ycau_xcri_xEin}
\end{figure}

\textbf{Caustic and critical curve.}
Caustics and critical curves are conveniently characterized by the image magnification,
\begin{equation}
    \mu = \left( \frac{\sin\beta}{\sin\theta} \frac{\dif \beta}{\dif \theta} \right)^{-1} ,
\end{equation}
which factorizes into a tangential and a radial piece,
\begin{equation}
    \mu_{\mathrm{t}} = \left( \frac{\sin\beta}{\sin\theta} \right)^{-1} ,
    \qquad
    \mu_{\mathrm{r}} = \left( \frac{\dif \beta}{\dif \theta} \right)^{-1} .
\end{equation}
Singularities of $\mu_{\mathrm{t}}$ and $\mu_{\mathrm{r}}$ define the tangential and radial critical curves in the image plane, respectively.
Their mappings into the source plane are the corresponding caustics.
In particular, the tangential critical curve (the Einstein ring) is located at $\thetaEin$, while the radial critical curve lies at $\thetacri$ with an associated source-plane radial caustic at $\betacau$.
Using the lens equation,
one obtains $\ycau\equiv\betacau\DOL/\mathcal{D}$ together with $\xcri\equiv\thetacri\DOL/\mathcal{D}$ and $\xEin\equiv\thetaEin\DOL/\mathcal{D}$ as functions of the parameter $\nu$.
These relations are shown in Fig.~\ref{fig:ycau_xcri_xEin}.
As evident in the figure, the radial caustic size $\ycau$ decreases monotonically with increasing $\nu$, and the caustic is absent for $\nu\ge 1$.
Since $\nu(t)$ is periodic, the radial caustic therefore pulsates in and out with the same period, a key driver of the periodic caustic-crossing lensing phenomenology.

\textbf{Event number.}
The number of lenses in a spherical shell of radius $\DOL$ and thickness $\dif\DOL$ is $n_a\,4\pi\DOL^2\dif\DOL$.
For a given $\DOL$, the number of sources that both satisfy $\DOS > \DOL^2/(\DOL-\rcS)$ and fall within the angular window $\beta_{\min}<\beta_{\src}<\beta_{\max}$ is
\begin{equation}
\begin{aligned}
&\int_{\DOL}^{Z} \dif\DOS \Big\{ n_b \,\Theta(\DOS-\frac{\DOL^2}{\DOL-\rcS}) \\
&\quad \times 2\pi\DOS^2 \; \max\left[0,\, \cos\beta_{\min} - \cos\beta_{\max} \right] \Big\} .
\end{aligned}
\end{equation}
Imposing the remaining condition $\DOL > \rcS $, the expected number of periodic caustic-crossing lensing events quoted in the main text is obtained.

Requiring the presence of a radial caustic ($\nu<1$) sets a lower bound on the mass of an oscillating boson star, $M>{\rc^2}/{(4D)}$.
Invoking the scale invariance of EKG equations in the small-amplitude regime, one finds $\tilde{r}_{\mathrm{c}} \propto \tilde{M}^{-1}$, which implies $\rc \propto M^{-1} m^{-2}$.
Consequently, the minimum mass capable of producing an oscillating caustic is
\begin{equation}
    M_{\min} \approx 1.6 \times 10^7 \Msun \left(\frac{1\Mpc}{D}\right)^{1/3} \left(\frac{10^{-20}\eV}{m}\right)^{4/3} ,
\end{equation}
For the representative geometry $Z=\DOS=2\DOL$ (so that $D=Z/4$), this becomes
\begin{equation}
    M_{\min} \approx 1.6 \times 10^7 \Msun \left(\frac{Z/4}{1\Mpc}\right)^{-1/3} \left(\frac{m}{10^{-20}\eV}\right)^{-4/3} .
\end{equation}

\end{document}